# Logistic Regression Analysis on the Dietary Behavior and the Risk of Nutritional Deficiency Dermatosis: The Case of Bicol Region, Philippines

**JOHN BEN S. TEMONES**
**johnben.temones@cbsua.edu.ph**

## ABSTRACT

This study explores the link between dietary behavior and the risk of nutritional deficiency dermatoses (NDD) in the Bicol region, where malnutrition remains a concern. Using regression analysis on FNRI data, it examines food purchase patterns, particularly riboflavin intake. Findings show an NDD risk prevalence of 15.75%, with Masbate and Camarines Sur contributing over half of cases. While rice (≤1590.93 g/day) and plant-based diets (523.30 g/day) are not rich in riboflavin, they still reduce NDD odds by 0.3% per gram. Riboflavin-rich foods like meat, eggs, and dairy lower risks by up to 3% per gram. The logistic regression model demonstrated strong performance (Nagelkerke = 0.765, accuracy = 94.1%, precision = 84.5%). Findings highlight the need for nutrition interventions, including enriched rice, better market access, and food diversity education to improve riboflavin intake and mitigate NDD risks.

Keywords: Nutritional deficiency dermatosis, riboflavin deficiency, food purchase patterns, machine learning, logistic regression analysis.

## INTRODUCTION

Nutritional deficiencies continue to be a pressing public health concern in many developing countries, particularly in regions with limited access to diverse food sources. In the Philippines, the Bicol region is known for its rich cultural heritage and agricultural productivity (per TourGuidePh), with recent reports from the DA Press Office (2022) that the region achieved a palay production growth of 4.01% and a corn output growth of 3.81% in 2021, yet it faces significant challenges related to malnutrition. According to World Bank (2021), many adults in the Philippines are affected by health conditions due to imbalanced nutrition, which leads to nutritional deficiencies and health complications. Among the various manifestations of nutrient deficiencies is the so-called nutritional deficiency dermatoses (NDD), which serve as an early indicator of broader malnutrition. According to Johnson (2024), skin conditions such as cheilosis, glossitis, and seborrheic dermatitis, often linked to riboflavin (vitamin B2) deficiency are prevalent in areas such as the Bicol region where dietary diversity is compromised by socioeconomic factors and limited access to nutrient-rich foods, as attested by Isaura, et.al (2022).

Though deemed as non-fatal, these conditions brought about by riboflavin deficiency which affects approximately 10-15% of the world population according to Marashly and Bohlega (2017), can impact a person's quality of life if not properly treated or prevented. Hence, understanding the dietary patterns that contribute to these deficiencies is crucial in addressing the public health challenges faced by Bicolano communities. Previous researches such as that of Li, et.al. (2022) has identified a strong correlation between dietary behavior such as food purchase patterns and the onset of deficiency-related conditions, but little has been done to quantify and predict these risks using advanced analytical techniques. Khoo (2023) showed that traditional methods for dietary assessment, such as 24-hour recalls and food diaries, often present challenges related to feasibility, accuracy, and cost.

This observation is reinforced by Ravelli and Schoeller (2020), who noted that self-reported dietary data are often subject to memory lapses and social desirability bias, particularly in low-literacy populations. Gibney et al. (2004) further warned that such biases can systematically distort the nutritional profiles of entire communities when used in epidemiological surveillance. In addition to methodological flaws, these traditional tools often fail to account for the dynamic and multi-dimensional nature of dietary behavior, especially in regions with seasonal variation in food availability.

The World Health Organization (2020) emphasized the need for developing more accurate and context-specific nutritional assessment tools that incorporate behavioral, economic, and environmental dimensions of food access. Such a need is even more pressing in the Bicol region, where agricultural productivity does not always translate to household dietary adequacy. Ruel-Bergeron et al. (2015) discussed the phenomenon of "hidden hunger," highlighting how energy sufficiency masks micronutrient deficiencies, especially in communities dependent on monotonous staple-based diets.

Moreover, studies by Drewnowski and Specter (2004) found that low-income populations tend to consume energy-dense but micronutrient-poor foods, driven primarily by economic constraints and food marketing strategies. Chakona and Shackleton (2019) support this by stating that in resource-constrained settings, economic factors are the strongest determinants of food choice, often outweighing nutritional knowledge or awareness. As such, understanding food purchase behavior becomes essential not just in capturing dietary trends, but in diagnosing underlying nutritional risks. Kennedy et al. (2007) and Arimond et al. (2010) both argued that dietary diversity scores can effectively predict micronutrient adequacy, but they also cautioned that these measures are context-sensitive and must be interpreted alongside socioeconomic data. In Bicol, for instance, access to diverse food groups is not merely a matter of availability but also one of affordability, preference, and cultural norms—all of which influence household purchasing decisions.

To bridge these complex layers of behavior and health, regression analysis offers a robust method for quantifying associations between food purchases and nutrient intake. Willett (2012) explains that regression allows for simultaneous adjustment of confounding variables, making it an ideal tool for public health nutrition where multiple factors are at play. Satija et al. (2015) showed that such methods could isolate the dietary determinants of specific nutritional outcomes like anemia, vitamin A deficiency, or stunting. In the same vein, Amugsi et al. (2013) employed regression models in sub-Saharan Africa to assess the influence of maternal education and household wealth on children's diet quality, revealing insights that were not apparent from descriptive statistics alone.

Despite the promising application of regression analysis in nutrition research, a significant gap remains in its use to model the direct relationship between food purchase behaviors and the risk of nutrient deficiency dermatoses in specific local contexts such as the Bicol region. Most studies, including those by Ivers and Cullen (2011) and Ferguson et al. (2015), have focused on general malnutrition or anthropometric outcomes, leaving a scarcity of research focused on dermatological markers of deficiency. Moreover, there is a lack of integrated models that combine actual purchase data with region-specific nutrient profiling to capture localized patterns of risk. This methodological and contextual gap underscores the need for a more nuanced analytical framework that reflects the lived dietary realities of vulnerable Filipino communities.

This study aims to fill this gap by employing regression analysis, integrating local food purchase patterns and nutritional secondary data from Food and Nutrition Research Institute (FNRI) to model the relationship between dietary behavior and the risk of NDD in the Bicol region, enhancing dietary assessment techniques which equates to greater objectivity and improved accuracy in predicting NDD. The findings are expected to contribute to the development of targeted nutritional interventions, providing policymakers with actionable insights into the dietary improvements needed to combat malnutrition. Additionally, the study serves as a model for using advanced data analysis techniques to address public health issues, with the potential for expansion into other regions and nutrient-related conditions.

## OBJECTIVES OF THE STUDY

The primary objective of this research is to analyze the relationship between dietary behavior and the risk of developing nutritional deficiency dermatoses (NDD) in the Bicol region, with a specific focus on riboflavin intake. By employing regression analysis, this study aimed at identifying key dietary patterns and food consumption habits that contribute to the prevalence of deficiency-related dermatoses, providing insights for targeted nutritional interventions and public health strategies. Specifically, this study sought to:

1) Determine the average daily purchase among Bicolanos of some specific food groups.
2) Determine the average daily riboflavin intake and the prevalence of nutrition deficiency dermatosis (NDD) in the Bicol region.
3) Describe the logistic regression model's overall fit and performance in predicting NDD.
4) Present the odds ratio for each food groups and its implication in predicting NDD.

## THEORETICAL FRAMEWORK OF THE STUDY

The theoretical framework for this study integrates nutritional epidemiology, dietary behavior theory, and machine learning-based health prediction to explain how dietary habits influence the risk of nutritional deficiency dermatoses (NDD) among Bicolanos. This framework is grounded in established theories and models that link nutrition, dietary choices, and disease risk.

***<u>Nutritional Epidemiology and the Diet-Disease Relationship.</u>*** This study builds on the Dietary Patterns and Disease Risk Model, which suggests that long-term dietary habits significantly influence the development of nutrient deficiencies and associated diseases. According to Willett (1998), dietary patterns—shaped by cultural, economic, and environmental factors—impact nutrient intake and the likelihood of deficiency-related conditions. In this study, riboflavin deficiency is the primary concern, as it is associated with conditions such as cheilosis, glossitis, and seborrheic dermatitis. The concept of nutritional adequacy is also essential to this framework, emphasizing that the adequacy of micronutrient intake (e.g., riboflavin) determines whether individuals experience deficiency symptoms. The research integrates this concept by assessing Bicolanos' food purchase patterns and determining whether they meet the Recommended Dietary Allowance (RDA) for riboflavin.

***<u>The Theory of Planned Behavior (TPB) in Dietary Choices.</u>*** This study aligns with Ajzen's (1991) Theory of Planned Behavior (TPB) to explain how individuals make dietary decisions that influence their nutritional status. TPB suggests that attitudes, subjective norms, and perceived behavioral control drive dietary behavior. In the context of this research,

- Attitudes: Bicolanos may prioritize staple foods (e.g., rice) over riboflavin-rich foods due to affordability and traditional dietary habits.
- Subjective norms: Cultural and societal influences, such as a preference for plant-based diets or economic constraints, shape food choices.
- Perceived behavioral control: Access to riboflavin-rich foods is affected by economic limitations, agricultural supply chains, and market accessibility.

By incorporating TPB, this study recognizes that food purchase patterns are not solely dictated by nutritional knowledge but are influenced by a complex set of social and economic factors.

***<u>Logistic Regression and Machine Learning in Nutritional Risk Prediction.</u>*** The use of logistic regression analysis aligns with predictive modeling approaches in public health, particularly in nutritional epidemiology. Hosmer & Lemeshow (2000) highlight logistic regression as a robust statistical method for estimating the probability of health outcomes based on predictor variables. In this study, logistic regression:

- Identifies significant predictors of NDD, such as daily riboflavin intake, food group purchases, and socio-economic factors.
- Provides odds ratios (OR) that quantify how food purchase patterns impact the risk of deficiency dermatoses.
- Helps develop a predictive model that can guide nutritional interventions and policy recommendations in the Bicol region.

Additionally, integrating machine learning approaches (e.g., feature selection, model evaluation) improves the accuracy of dietary risk assessment, overcoming the limitations of traditional survey-based dietary assessments.

## METHODOLOGY

This study employed a cross-sectional research design to investigate the relationship between dietary behavior and the risk of nutritional deficiency dermatoses among Bicolano households. The research primarily focused on analyzing food purchase patterns and its association with the prevalence of deficiency-related skin conditions such as cheilosis, glossitis, and seborrheic dermatitis as determined by estimated riboflavin intake. The methodology of this research is comprehensively discussed below.

**Data Collection, Sampling, and Descriptive Statistics Used**

A secondary data was used in this study, particularly from the Expanded National Nutrition Survey (ENNS): Dietary Component Individual Food Consumption 2018, 2019, and 2021 of the Food and Nutrition Research Institute, promptly requested through their eNutrition website. ENNS utilized a rolling sample design in which 117 Philippine Statistics Authority (PSA) domains of provinces and highly urbanized cities (HUCs) were grouped into 24 replicates having similar characteristics utilizing data from the 2010 Census of Population and Housing (CPH). Eight replicates consisting of 40 domains were independently allocated each year to generate national-level estimates. For the regional estimates, the total provinces/HUCs covered in the three periods of the ENNS were cumulated to come up with the regional level data of the different nutrition and health indicators collected in the survey. ENNS furthermore employed a two non-consecutive day 24-hour food recall to assess the food intake of individuals via their daily purchase patterns of specific food groups, some of which were considered as predictors of riboflavin intake and risk of NDD in the study namely; (a) cereals and cereal products, (b) starchy roots and tubers, (c) sugar and syrups, (d) dried beans, nuts, and seeds, (e) vegetables, (f) fruits, (g) fish and fish products, (h) meat and meat products, (i) poultry, (j) eggs, (k) milk and milk products, (l) fats and oil, and (m) beverages. For this study, a total of 724 individuals from the Bicol Region were extracted from the ENNS dataset, spread among its 6 provinces below:

Table 1. Sample distribution among Bicol's Provinces.

| Province | No. of Samples |
|---|---|
| Albay | 169 |
| Camarines Norte | 75 |
| Camarines Sur | 210 |
| Catanduanes | 40 |
| Masbate | 129 |
| Sorsogon | 101 |
| TOTAL | 724 |

Moreover, mean was used to give a holistic view of the daily purchase patterns among Bicolanos since no significant outliers were found in each individual food groups after employing Grubb's test.

**Riboflavin Assessment and Prevalence of NDD**

The Individual Dietary Evaluation System (IDES) developed by DOST-FNRI was used to estimate the amount of food, including energy and nutrient content of foods consumed by each individual, hence the ENNS dataset already contains the daily riboflavin intake for each individual among other nutrients. It was mentioned in the study of Poonam (2022) that on average, adults need between 1.3 and 1.6 milligrams (mg) of riboflavin every day to avoid a deficiency. This literature was used to assess potential risk of the prevalence of NDD among Bicolanos, particularly by noting individuals with riboflavin intake ≤ 1.3 mg as riboflavin deficient (hence potentially at risk of NDD) coded as “1” and coded as “0” otherwise. Prevalence, according to National Institute of Mental Health (NIMH), is the number of people in the sample with the characteristic of interest, divided by the total number of people in the sample. Therefore, we identify the prevalence of risk of NDD by following the formula below:

$$\text{Prevalence} = \frac{\text{No. of Bicolanos Potentially at Risk of NDD}}{\text{Overall Sample Size}} \times 100$$

## Logistic Regression Analysis

To analyze the association between dietary behavior and the risk of nutritional deficiency dermatoses, logistic regression was applied. Logistic regression, among other machine learning methods, was strongly considered in this study since the target variable is categorical and dichotomous, representing the potential presence or absence of risk of dermatoses, while the independent variables (predictors) are the purchase patterns from the specified food groups. The model allowed for the estimation of odds ratios, indicating the likelihood of developing dermatoses based on their dietary behavior. For instance, the model examined how the frequency of purchasing certain food groups influenced the probability of riboflavin deficiency and developing skin-related conditions. Results were reported with p-values and confidence intervals to provide clarity on the strength and significance of the relationships between dietary factors and dermatoses. Sensitivity analyses were performed to assess the robustness of the results, and multicollinearity was checked to ensure the reliability of the regression model.

This methodological approach provides a comprehensive framework for understanding the dietary behaviors of Bicolano households and their potential link to NDD, enabling the study to draw meaningful conclusions regarding dietary interventions and public health strategies aimed at reducing the prevalence of dermatoses in the region.

# RESULTS AND DISCUSSION

This chapter presents the findings of the study, analyzing the relationship between dietary behavior and the risk of developing nutritional deficiency dermatoses in the Bicol region of the Philippines. Using logistic regression, the analysis evaluated how various dietary factors, particularly riboflavin intake, influence the likelihood of developing skin-related conditions such as cheilosis, glossitis, and seborrheic dermatitis. The discussion integrates these results with existing literature, highlighting the significance of dietary patterns and nutrient intake in preventing deficiency-related conditions. Additionally, potential implications for public health interventions in the region are considered, with a focus on improving nutrition through targeted dietary changes.

## Average Daily Purchase of Specific Food Groups Among Bicolanos

This subsection examines the average daily purchase of specific food groups among Bicolano households, providing insights into their dietary behavior and nutritional intake. Understanding these purchasing trends is crucial for assessing how daily food choices may contribute to the risk of nutritional deficiency dermatoses in the region. This analysis also highlights the accessibility and affordability of these specific food groups in local markets. Figure 1 shows the average daily purchase of these food groups.

| | Valid | Missing | Mean | Std. Deviation |
|---|---|---|---|---|
| Vegetables | 724 | 0 | 523.300 | 560.373 |
| Fruits | 724 | 0 | 139.294 | 401.180 |
| Fish and Fish Products | 724 | 0 | 428.034 | 508.987 |
| Meat and Meat Products | 724 | 0 | 160.470 | 295.644 |
| Poultry | 724 | 0 | 67.370 | 219.332 |
| Eggs | 724 | 0 | 53.525 | 100.424 |
| Milk and Milk Products | 724 | 0 | 194.816 | 730.881 |
| Fats and Oils | 724 | 0 | 91.527 | 160.084 |
| Beverages | 724 | 0 | 57.681 | 122.207 |
| Cereals and Cereal Products | 724 | 0 | 1590.930 | 874.365 |
| Starchy Roots and Tubers | 724 | 0 | 65.702 | 294.393 |
| Sugar and Syrups | 724 | 0 | 64.137 | 78.535 |
| Dried Beans, Nuts, and Seeds | 724 | 0 | 24.952 | 70.454 |

Figure 1. Average Daily Purchase (in grams) of Bicolanos

As gleaned from the table, the main staple of Bicolanos revolve around cereals (1590.93 g/day), which according to FNRI include rice (ordinary, special, and glutinous), other rice products such as rice noodles (bihon), rice cakes (puto), biko, suman, arrozcaldo, champorado, milled corn, corn on a cob, other corn products like cornstarch, corn pudding (maja blanca), popcorn, corn chips, pandesal, bread, cookies/biscuits, cakes/pastries, noodles, flour, and others. Rice alone, which is obviously common in Filipino tables, constitutes about 87.52% of this daily purchase (at 1392.31 g/day) according to the data. This implies that Bicolanos may generally have insufficient daily intake of riboflavin, taking into account that the most common type of rice consumed by Asians according to Goh (2018) is the medium-grain rice, which only contains 0.01mg of riboflavin per 100g according to the literature from Fit Audit (2024). Consumption of enriched rice, however, can be a beneficial option for those seeking to increase their intake of riboflavin as it is usually fortified with nutrients to enhance its nutritional value according to the National Institute of Health (2024). In connection to this, it is important to note that cooking methods can impact the riboflavin content in rice. Riboflavin is water-soluble, meaning a portion of it can be lost during cooking, particularly if rice is rinsed prior to cooking. To minimize nutrient loss, it is advisable not to rinse rice after cooking per UMass Amherst. (n.d.).

It is also worth noting from figure 1 that the second most prioritized food in terms of daily purchase amongst Bicolanos are vegetables, with 523.3 g/day. Vegetables do provide some riboflavin, but they are generally not the richest sources compared to other notable food groups such as dairy products (194.816 g/day), eggs (53.525 g/day), and meat (160.47 g/day) which are considered to be the most abundant sources of riboflavin according to Poonam (2022) but is alarmingly at the bottom of the food priority of Bicolanos. One of the probable reasons behind this data is the primary concern for health for many individuals. For instance, according to Pickles (2024) high consumption of red and processed meats has been linked to numerous health issues, including heart disease, diabetes, and various cancers thus people selecting a plant-based diet. Moreover, according to Scott-Reid (2021) economic factors such as rising costs associated with sustainably and ethically sourced meat and dairy often push consumers toward cheaper, plant-based alternatives. The perception of these products being more expensive creates a barrier for many consumers, making it more appealing to choose plant-based options that fit within their budget, specifically that a typical Bicolano family only spends an average of Php 202,620.00 yearly (roughly Php 555.12 per day) according to the Family Income and Expenditure Survey of Philippine Statistics Authority (PSA) as reported by Mapa in 2024.

While one may argue that these amounts collectively contribute to daily riboflavin needs, it is important to note in this discussion how instrumental it is to include a variety of foods in your diet to ensure adequate intake of riboflavin.

## Average Daily Riboflavin Intake and the Prevalence of NDD in Bicol

According to the FNRI data, Bicolanos have an average daily riboflavin intake of 2.96 mg/day, more than twice the average from Poonam's (2022) study to avoid riboflavin deficiency. This is despite the contrasting data on the food purchase pattern of Bicolanos which largely revolve on cereals (particularly rice) and vegetables. This may be attributed to some foods that are not typically recognized as high in riboflavin but can still contribute to overall intake. For example, dried beans, nuts, and seeds can provide moderate amounts of riboflavin, with an ounce providing 23% of the riboflavin you need in a day and as seen from figure 1, Bicolanos buy 24.952 grams of these daily which is about 0.88 ounce. The table below shows the number of Bicolanos with a potential risk of NDD.

Table 2. Riboflavin Deficiency Amongst Provinces of Bicol

| Province | Sample Size | Riboflavin Deficient | Percentage |
|---|---|---|---|
| Albay | 169 | 18 | 10.65 |
| Camarines Norte | 75 | 10 | 13.33 |
| Camarines Sur | 210 | 28 | 13.33 |
| Catanduanes | 40 | 11 | 27.50 |
| Masbate | 129 | 33 | 25.58 |
| Sorsogon | 101 | 14 | 13.86 |
| **BICOL REGION** | **724** | **114** | **15.75** |

It can be gleaned from table 2 that the 15.75% of Bicolanos are riboflavin deficient, and hence is prone to be having NDD. This number is at most 6% higher than the approximation in the study of Marashly and Bohlega (2017), though it may be more prevalent than currently acknowledged. This underscores a pressing public health concern and suggests potential gaps in local nutritional adequacy. These limitations often lead to reliance on riboflavin-less staples. For instance, the elevated prevalence indicates broader dietary challenges faced by the population, highlighted by the data that Bicolanos tend to purchase foods according to their customs (rice dependence), other health beliefs (plant-based diet), and their purchasing capacity. Addressing these disparities requires targeted nutrition interventions, agricultural boost, and public health strategies to improve access to riboflavin-rich foods, emphasizing the need for policies that support dietary diversity and food fortification in the region.

Notably, Camarines Sur and Masbate collectively account for over half of the NDD cases recorded in the Bicol region, warranting closer examination of their shared vulnerabilities. Several interrelated factors may explain this trend. First, both provinces are marked by elevated poverty incidence and relatively low household incomes, as documented by the Philippine Statistics Authority (2024), which limits access to nutrient-dense foods such as eggs, milk, and meat—key sources of riboflavin. Drewnowski and Specter (2004) argue that in low-income settings, households tend to prioritize energy-dense but micronutrient-poor food items, a behavior driven by economic necessity rather than nutritional adequacy.

Masbate's geographic isolation as an island province may further exacerbate the situation by restricting access to diverse food markets and perishable nutrient-rich goods. Ezeh et al. (2020) note that communities in geographically disadvantaged areas often face logistical and seasonal barriers that contribute to lower dietary diversity and increased risk of nutrient deficiencies. Camarines Sur, while agriculturally productive, experiences significant rural inequalities. Isaura et al. (2022) emphasized that agricultural abundance does not necessarily translate to household-level nutritional adequacy, particularly when food production is geared towards income generation rather than local consumption.

Cultural dietary norms may also reinforce these deficiencies. Bicolano households tend to rely heavily on rice and plant-based meals, which, though affordable and satiating, are relatively poor in riboflavin. As Chakona and Shackleton (2019) observed, economic constraints often outweigh nutritional knowledge, leading households to opt for foods that align with their financial capacities rather than those that meet micronutrient requirements. This dietary monotony, coupled with limited purchasing power and market access, may contribute to the elevated prevalence of NDD in these provinces. Thus, targeted nutrition programs in Masbate and Camarines Sur should consider both structural barriers and behavioral patterns, focusing on food fortification, improving accessibility of riboflavin-rich foods, and promoting culturally sensitive dietary education.

## Logistic Regression Analysis on Food Purchase Pattern and NDD Prevalence

The logistic regression model was used to predict the likelihood of developing NDD based on the food purchase patterns of Bicolanos on a daily basis. The model was deemed to be fit since our target variable is binary, where 1 indicates potential risk of dermatoses and 0 indicates the opposite. The model demonstrated a good fit (Nagelkerke $R^2 = 0.765$, $p < 0.001$), explaining 76.5% of the variance in the risk of NDD. Additionally, the model performed well in predicting the NDD, with overall accuracy of 94.061% as shown in the confusion matrix below. This means that for every 100 predictions that the model makes, it will accurately predict at most 95 of the cases correctly.

Confusion matrix

| Observed | Predicted 0 | Predicted 1 | % Correct |
|---|---|---|---|
| 0 | 594 | 16 | 97.377 |
| 1 | 27 | 87 | 76.316 |
| Overall % Correct | | | 94.061 |

*Note*. The cut-off value is set to 0.5

Figure 2. Confusion Matrix of the Model

While accuracy is a useful metric, it does not always provide a complete picture, especially if the dataset is imbalanced where some classes are much more common than others. This is in fact the case in this study, where there are way lesser instances of Bicolanos at risk of NDD than those who are not. Figure 3 shows the other performance metrics of the model. It can be gleaned from the figure that the overall precision of the model is 84.5%, which means that when the model makes a prediction for the existence of NDD, the model would predict at most 85 NDD cases out of 100 predictions correctly. When it comes to predicting actual positive cases of NDD as opposed to those classified as false negatives (sensitivity), the model predicts 76.3% of the cases correctly. On the other end, when the model predicts actual negative cases of NDD as opposed to those classified as false positives (specificity), the model predicts 97.4% of such cases correctly. Moreover, variance inflation factors (VIFs) of the predictors were all found to be below 2, indicating no multicollinearity issues among the predictor variables.

Performance metrics ▼

| | Value |
|---|---|
| Accuracy | 0.941 |
| Sensitivity | 0.763 |
| Specificity | 0.974 |
| Precision | 0.845 |

Figure 3. Performance Metrics of the Model

From table 3, it can be gleaned that cereals and cereal products is a significant predictor (OR=0.997, $p<.001$) of NDD, which means that for every additional gram of purchase of these products, the odds of developing NDD decrease by 0.3%. The case is similar to that of vegetables, even when they are not good sources of riboflavin. Notable sources of riboflavin, namely meat and meat products (OR=0.993, $p<0.001$), milk and milk products (OR=0.992, $p<0.001$), poultry (OR=0.981, $p<0.001$) and eggs (OR=0.970, $p<0.001$) were typically found to be significant predictors of NDD among Bicolanos, which reduces the odds of developing NDD by 0.7%, 0.8%, 1.9%, and 3.0% respectively for every additional gram of purchase. Purchase of fish also significantly reduce the odds of NDD by 0.4%, and surprisingly, every additional gram of purchase of beverages can greatly contribute in significantly decreasing the odds of NDD by 1%. Conversely, though the odds ratio of purchasing sugar and syrups, dried beans, nuts, and seeds, fruits, and fats and oil show evidence of reducing the odds of NDD for every additional gram of purchase, the model showed that these are not significant. This means that purchasing these products, though some have reasonable of riboflavin value based from prior studies and literatures such as that from WebMD (Poonam, 2024), will not significantly lessen the chance of developing NDD among Bicolanos. On the other end, though every additional gram of purchase of starchy roots and tubers was found to increase the chance of having NDD by 0.1% (OR=1.001), this does not pose significant result (p=0.056), hence showing evidence of no NDD harm to purchase these products.

Table 3. Odds Ratio of the Food Groups in Predicting NDD

| Predictor | OR | p-value |
|---|---|---|
| Cereals and Cereal Products | 0.997 | <.001 |
| Starchy Roots and Tubers | 1.001 | 0.056 |
| Sugar and Syrups | 0.998 | 0.700 |
| Dried Beans, Nuts, and Seeds | 0.984 | 0.061 |
| Vegetables | 0.997 | <.001 |
| Fruits | 0.999 | 0.544 |
| Fish and Fish Products | 0.996 | <.001 |
| Meat and Meat Products | 0.993 | <.001 |
| Poultry | 0.981 | <.001 |
| Eggs | 0.970 | <.001 |
| Milk and Milk Products | 0.992 | <.001 |
| Fats and Oils | 0.995 | 0.086 |
| Beverages | 0.990 | 0.013 |

These findings suggest that interventions aimed at increasing riboflavin intake could reduce the risk of NDD in the Bicol Region. For instance, taking into account the obsession of Bicolanos (Filipinos, in general) with rice, government units and other research bodies may look into possible development and introduction to enriched varieties of rice, which can be equipped with additional nutrients such as riboflavin since rice is found to significantly reduce the odds of NDD. Moreover, since Bicolanos have low expenditure resulting from relatively low annual income compared to other regions as shown in the PSA data, government could provide assistance to Bicolanos by improving access to markets with products rich in riboflavin to ensure more room for spending on diverse food groups, particularly in areas with limited food diversity.

## CONCLUSIONS AND RECOMMENDATIONS

This study explored the relationship between dietary behavior and the risk of nutritional deficiency dermatoses (NDD) in the Bicol region of the Philippines by employing logistic regression analysis in an existing data from the Food and Nutrition Research Institute (FNRI) on food purchase patterns of Bicolanos on specific food groups and riboflavin intake among Bicolano households, which is a key contributor in the development of dermatosis. Findings revealed that most Bicolanos spend their money to buy cereals and cereal products (1590.93 g/day) and vegetables (523.30 g/day). The prevalence of nutritional deficiency dermatosis risk in Bicolanos is at 15.75%, with Masbate and Camarines Sur collectively contributing more than half of these cases. The logistic regression model showed good fit, with significant Nagelkerke $R^2$ value of 0.765, explaining 76.5% of the variance in the risk of NDD. Moreover, metrics showed promising performance of the model, with overall accuracy of 94.1%, 84.5% precision, 76.3% sensitivity, and 97.4% specificity, with variance inflation factors (VIFs) all below 2, indicating no multicollinearity issues among the food groups. Cereals and vegetables were found to significantly reduce the odds of NDD by 0.3% for every additional gram of purchase even when they are not notable sources of riboflavin according to existing literatures. Typical sources of riboflavin such as meat and meat products (OR=0.993, $p<0.001$), milk and milk products (OR=0.992, $p<0.001$), poultry (OR=0.981, $p<0.001$) and eggs (OR=0.970, $p<0.001$) were found to be significant predictors of NDD among Bicolanos, which reduces the odds of developing NDD by 0.7%, 0.8%, 1.9%, and 3.0% respectively for every additional gram of purchase. Purchase of fish also significantly reduce the odds of NDD by 0.4%, while every additional gram of purchase of beverages can also greatly contribute in significantly decreasing the odds of NDD by 1%. On the other hand, purchasing sugar and syrups, dried beans, nuts, and seeds, fruits, and fats and oil show evidence of reducing the odds of NDD for every additional gram of purchase but are not significant; and though every additional gram of purchase of starchy roots and tubers were found to increase the chance of having NDD by 0.1% (OR=1.001), this does not pose significant result ($p=0.056$).

This study concludes that even for food groups with little known riboflavin content such as rice, vegetables, and beverages, sufficient purchase of these products can significantly decrease the chance of acquiring NDD, and can further complement the riboflavin sufficiency of noted riboflavin-rich foods such as meats, dairy products, and eggs. Purchasing riboflavin-rich foods such as dried nuts and seeds does not mean less chance of developing NDD among Bicolanos. On the other end, purchasing food groups which show evidence of increasing the odds of riboflavin deficiency such as starchy roots do not mean harm in acquiring NDD. Taking into account the food purchase pattern and the economic background of Bicolanos, this study strongly suggests some public health interventions aimed at increasing riboflavin intake to significantly reduce the incidence of nutritional deficiency dermatoses in the Bicol region. First, government may look into possible efforts in developing enriched varieties of rice, with emphasis in incorporating riboflavin in these varieties. Campaign programs aimed at introducing this and other already known riboflavin rich foods in the public with emphasis to food diversity is also highly encouraged. Targeted dietary programs focusing on increasing the availability of riboflavin-rich foods, such as dairy products and meat products, may also be particularly effective. Moreover, since Bicolanos are known to have low annual income hence low expenditure capacity, government could provide assistance to Bicolanos by improving access to markets with products rich in riboflavin to ensure more room for spending on diverse food groups, particularly in areas with limited food diversity. Controlling the prices of these products is also highly encouraged.

**Future research direction.** The interaction effects between food groups and cost analysis of proposed intervention in addressing nutritional deficiency dermatosis, which were considered as limitations of this study, may be explored in future researches. Deeper analysis of why food groups known to have little to none riboflavin such as beverages have a significant impact in decreasing chances of NDD may also be explored.

**Acknowledgement.** The researcher would like to extend its heartfelt gratitude to the Food and Nutrition Research Institute (FNRI) of the Department of Science and Technology (DOST) for providing access to their data from the Expanded National Nutrition Survey: Dietary Component, Individual Food Consumption for years 2018, 2019, & 2021 which paved the way for this research. Most of all, to CBSUA's SUC President IV, Dr. Alberto N. Naperi, for his unwavering support to the researcher in presenting this research to the 2024 Bicol Pulse: Health Research Competition organized by DOST Bicol Consortium for Health Research and Development (BCHRD) to which it won as the 3rd best paper.